\documentclass[review,1p,times,sort&compress]{elsarticle}

\usepackage{amssymb}

\journal{Physica A}

	\usepackage{fancyheadings}
\renewcommand{\thispagestyle}[1]{} 

\begin{document}

\begin{frontmatter}



\title{Ferrimagnetism in the Heisenberg-Ising Bilayer with Magnetically Non-equivalent Planes}

\author[]{Tadeusz Balcerzak}
\author[]{Karol Sza\l{}owski\corref{label0}}
\ead{kszalowski@uni.lodz.pl}
\cortext[label0]{corresponding author}

\address{Department of Solid State Physics, Faculty of Physics and Applied Informatics, University of
{\L}\'{o}d\'{z}, ul. Pomorska 149/153, 90-236 {\L}\'{o}d\'{z},
Poland}

\begin{abstract}
The Pair Approximation method is applied to the antiferromagnetic Heisenberg-Ising spin-1/2 bilayer with a simple cubic crystalline structure. The method allows for self-consistent calculations of thermodynamic quantities, based on the determination of Gibbs free-energy. In the paper the phase diagrams and planar magnetizations are calculated for the system in question. Special attention is paid to the case of magnetically non-equivalent planes, one of which is additionally randomly diluted. The occurrence of a compensation phenomenon is found and the compensation temperature is discussed for such a system. The characteristic concentration of magnetic atoms $p^{\star}$, below which the compensation phenomenon vanishes, is examined as a function of the Hamiltonian parameters.

\end{abstract}

\begin{keyword}
Ising-Heisenberg model \sep aniferromagnetic bilayer \sep  critical temperature \sep compensation temperature \sep planar magnetization
\end{keyword}

\end{frontmatter}

\newpage

\section{Introduction}

Magnetic bilayers have attracted considerable attention in recent literature \cite{Hansen1,Jascur1,Horiguchi1,Lipowski1,Li1,Araujo1,Wiatrowski1,Mirza1,Spirin1,Ghaemi1,Monroe1,Kapor1,Saber1,Qiu1,Diep1,Balcerzak1,Szalowski1}. Such systems are interesting both from experimental and theoretical point of view since they bridge the gap between the two- and three-dimensional magnets. The other reason is that the low-dimensional magnets have gained an immense applicational potential
\cite{Zhitomirsky1,Honecker1,Schmidt1,Honecker2,Pereira1,Ribeiro1}. In particular, it is worth mentioning that giant magnetoresistance phenomenon is based on the antiferromagnetic coupling between ferromagnetically ordered magnetic planes \cite{Grunberg1,Fert1}.\\

Until now, numerous works have concentrated mainly on the bilayer Ising ferromagnets \cite{Hansen1,Horiguchi1,Lipowski1,Li1,Mirza1,Ghaemi1,Monroe1}. Some papers have extended the studies to the interesting case of the spins larger than 1/2 \cite{Horiguchi1,Kapor1,Balcerzak2}, as well as to different spins in both magnetic planes \cite{Jascur1,Balcerzak3,Albayrak1}. In several works the dilution \cite{Jascur1,Ainane1}, amorphous structure \cite{Bengrine1,Timonin1}, and the case of unequal coupling strengths in each magnetic plane \cite{Oitmaa1} have been considered. Moreover, the anisotropic Heisenberg bilayers have been studied, for instance, in Refs. \cite{Balcerzak1, Szalowski1,Szalowski2}.\\

As far as the ferromagnetic bilayers with antiferromagnetic couplings are concerned, there are only very few theoretical works dealing with this problem \cite{Kim1,Sankowski1,Szalowski3,Szalowski4} or a similar case of multilayers \cite{Deviren1,Stier1}, to the best of our knowledge. Let us mention the existence of some recent interesting experimental realizations of the systems in question based on semiconducting compounds, both with and without dilution \cite{Furdyna1,Furdyna2,Furdyna3,Story2,Story3,Story1}. In order to fill the gap, the present paper aims at examining the thermodynamics of the anisotropic Heisenberg-Ising bilayer with antiferromagnetic interplanar coupling. Within both magnetic planes forming the bilayer, the intraplanar couplings are anisotropic and ferromagnetic. For the sake of generality, both planes are assumed to be magnetically non-equivalent, thus characterized by the exchange integrals of different strength and anisotropy in the spin space. Moreover, we assume that the plane with stronger magnetic couplings is randomly diluted. Such a model introduces the possibility to examine the occurrence of the compensation point in the presence of the antiferromagnetic interplanar interactions and its dependence on the Hamiltonian parameters. Therefore, the aim of the paper is to study the phase transition and compensation temperatures, as well as the planar magnetizations of the bilayer as a function of magnetic anisotropy, antiferromagnetic coupling strength and dilution.\\

To achieve the aim we developed the cluster variational method in the Pair Approximation (PA) and we adopted it for the system in question. The PA method has already been successfully applied in several cases, including studies of ferromagnetic Heisenberg bilayers with the spin $S=1/2$ \cite{Balcerzak1, Szalowski1,Szalowski2} and the Blume-Emery-Griffiths model \cite{Tucker1}. The approach is considerably more sophisticated and accurate than Molecular Field Approximation since it takes into account the correlations between pairs of spins. Let us note that the PA method allows for a complete construction of the thermodynamic description based on the expression for Gibbs free energy for the system.\\

The paper is organized as follows: in the theoretical section (2) we develop the PA method for the anisotropic bilayer with antiferromagnetic interplanar couplings and dilution. In the third section (3) we present and discuss the results of numerical calculations for the phase transition and compensation temperatures as well as the planar magnetizations. In the last section (4) the final remarks and conclusion are drawn.\\

\section{Theoretical model}
\subsection{The Gibbs free-energy}

\begin{figure}
\includegraphics[scale=0.25]{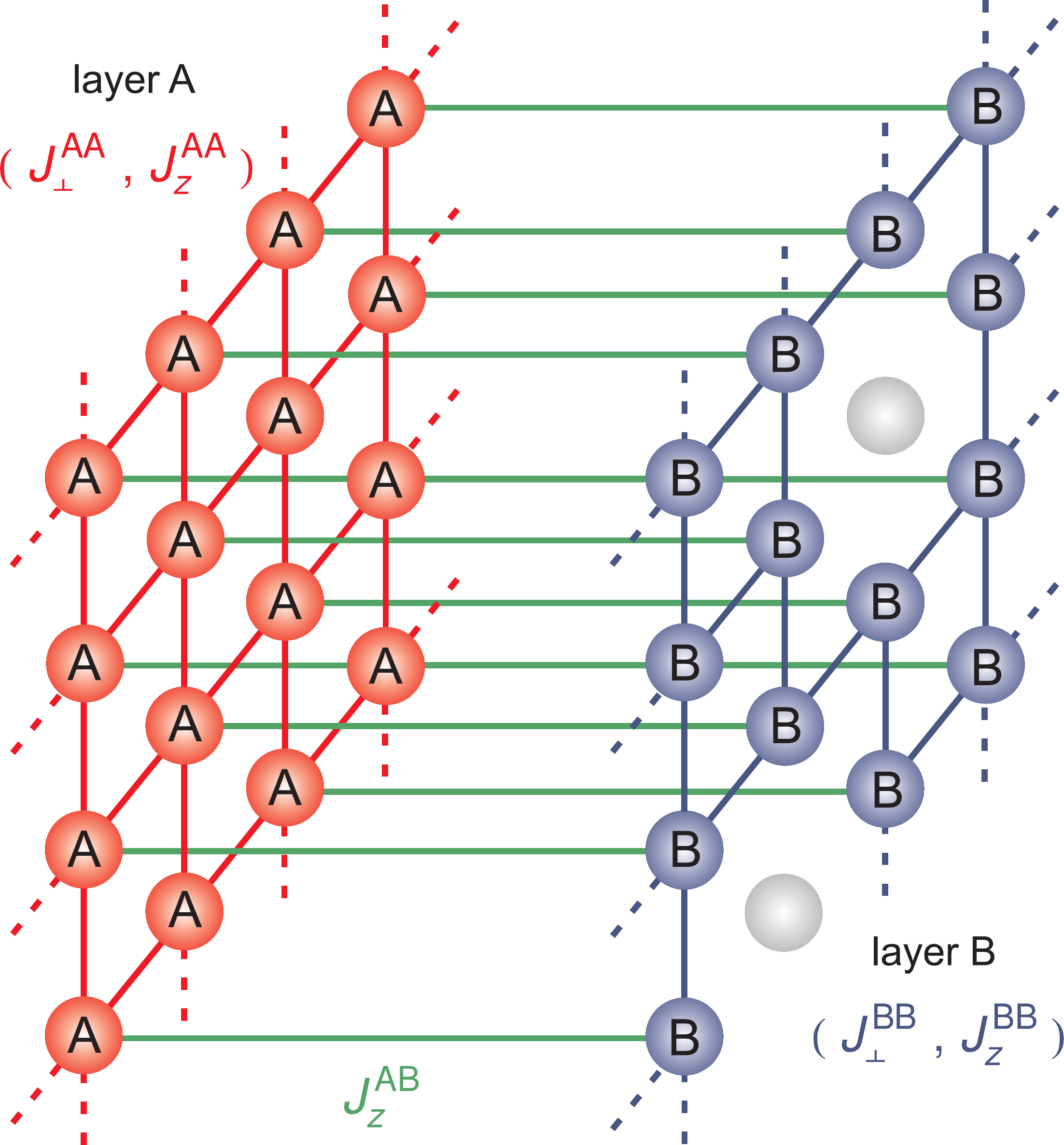}
\caption{\label{fig:fig1}A schematic view of the bilayer composed of two layers, $A$ and $B$. The intraplanar couplings are $J^{AA}_{x}=J^{AA}_{y}=J^{AA}_{\perp}$, $J^{AA}_{z}$ and $J^{BB}_{x}=J^{BB}_{y}=J^{BB}_{\perp}$, $J^{BB}_{z}$, respectively. The interplanar coupling is $J^{AB}_{z}$. The plane $B$ is randomly diluted.}
\end{figure}

The schematic view of the bilayer possessing simple cubic (s.c.) crystalline structure is illustrated in Fig.1.
The Hamiltonian of such a system can be written in the form of:
\begin{eqnarray}
\mathcal{H}&=&-\sum_{\left\langle i\in A,j\in A \right\rangle}^{}{\left[J_{\perp}^{AA}\left(S_{x}^{i}S_{x}^{j}+S_{y}^{i}S_{y}^{j}\right)+J_{z}^{AA}S_{z}^{i}S_{z}^{j}\right]}-\sum_{\left\langle i\in A,j\in B \right\rangle}^{}{J_{z}^{AB}S_{z}^{i}S_{z}^{j}\xi_j}\nonumber\\
&&-\sum_{\left\langle i\in B,j\in B \right\rangle}^{}{\left[J_{\perp}^{BB}\left(S_{x}^{i}S_{x}^{j}+S_{y}^{i}S_{y}^{j}\right)+J_{z}^{BB}S_{z}^{i}S_{z}^{j}\right]\xi_i\xi_j}-h\sum_{i\in A}^{}{S^{i}_{z}}-h\sum_{j\in B}^{}{S^{j}_{z}\xi_j}.
\label{eq1}
\end{eqnarray}
In the Hamiltonian, $J^{AA}_{\perp, z}$ and $J^{BB}_{\perp, z}$ are the anisotropic exchange integrals in the plane
$A$ and $B$ respectively, and $J^{AB}_{z}<0$ is the antiferromagnetic, Ising-type interaction between these planes. $h$ stands for the external magnetic field. We consider the case when $0\leq J^{\gamma \gamma}_{\perp} \le J^{\gamma \gamma}_{z}$ ($\gamma=A,B$), which covers all the intermediate situations between the pure Ising ($J^{\gamma \gamma}_{\perp}=0$) and the isotropic Heisenberg ($J^{\gamma \gamma}_{\perp}=J^{\gamma \gamma}_{z}$) couplings within $A$ and $B$ planes. In order to avoid a possible non-classical ground state, the interaction between $A$ and $B$ planes is always taken in the Ising form, i.e., $J^{AB}_{\perp}=0$. The site occupation operators $\xi_j=0,1$ describe the fact that the plane $B$ is randomly diluted. The configurational mean value of $\xi_j$, $\left< \xi_j \right> =p$ gives the concentration of magnetic atoms in that plane.\\ 

Following the method described in Refs. \cite{Balcerzak1, Szalowski1,Szalowski2}, Gibbs energy per site can be found. For the bilayer system with s.c. structure, with plane $B$ diluted with concentration $p$, Gibbs energy can be written in the form of:
\begin{equation}
G=G^{AA}+\frac{p}{2}G^{AB}+p^{2}G^{BB}-\frac{1}{2}\left(3+p\right)G^{A}-2p^{2}G^{B},
\label{eq2}
\end{equation}
where the single-site Gibbs energies are given by:
\begin{equation}
G^{\gamma}=-k_{\rm B}T \,\ln\left\{2\cosh\left[\frac{\beta}{2}\left(\Lambda^{\gamma}+h\right)\right]\right\} 
\label{eq3}
\end{equation}
for $\gamma =A,B$.\\
The pair Gibbs energies are given in the form of:
\begin{equation}
G^{\gamma \delta}=-k_{\rm B}T \,\ln\left\{2\exp\left(\frac{\beta J^{\gamma \delta}_{z}}{4}\right)\cosh\left[\beta\left(\Lambda^{\gamma \delta}+h\right)\right]+2\exp\left(-\frac{\beta J^{\gamma \delta}_{z}}{4}\right)\cosh\left[\frac{\beta}{2}\sqrt{\left(\Delta^{\gamma \delta}\right)^2+\left(J_{\perp}^{\gamma \delta}\right)^2}\,\right]\right\} 
\label{eq4}.
\end{equation}
where $\gamma=A,B$ and $\delta=A,B$.\\
The molecular fields $\Lambda^{\gamma}$, $\Lambda^{\gamma \delta}$ and $\Delta^{\gamma \delta}$ occurring in eqs. (\ref{eq3}-\ref{eq4}) can be expressed by four variational parameters $\lambda^{AA}$, $\lambda^{BB}$, $\lambda^{AB}$ and $\lambda^{BA}$, namely:
\begin{eqnarray}
\Lambda^{A}&=&4\lambda^{AA}+p\lambda^{AB}\nonumber\\
\Lambda^{B}&=&4p\lambda^{BB}+\lambda^{BA}\nonumber\\
\Lambda^{AA}&=&3\lambda^{AA}+p\lambda^{AB}\nonumber\\
\Lambda^{BB}&=&\left(4p-1\right)\lambda^{BB}+\lambda^{BA}\nonumber\\
\Lambda^{AB}&=&2\left(\lambda^{AA}+p\lambda^{BB}\right)+\frac{1}{2}\left(p-1\right)\lambda^{AB}\nonumber\\
\Delta^{AA}&=&0 \nonumber\\
\Delta^{BB}&=&0 \nonumber\\
\Delta^{AB}&=&4\left(\lambda^{AA}-p\lambda^{BB}\right)+\left(p-1\right)\lambda^{AB}
\label{eq5}
\end{eqnarray}
where $\lambda^{\gamma \delta}$ is the field acting on a spin in the plane $\gamma$ and originating from the spins in the planes $\delta$.\\
These four fields can be found from the variational principle for the Gibbs free energy:
\begin{equation}
\frac{\partial G}{\partial \lambda^{\gamma \delta}}=0
\label{eq6}
\end{equation}
As a result, $\lambda^{\gamma \delta}$ can be determined from the solutions of the following four variational equations:
\begin{eqnarray}
\tanh\left[\frac{\beta}{2}\left(\Lambda^{A}+h\right)\right]&=&\frac{\exp\left(\frac{\beta J^{AA}_{z}}{4}\right)\sinh\left[\beta\left(\Lambda^{AA}+h\right)\right]}{\exp\left(\frac{\beta J^{AA}_{z}}{4}\right)\cosh\left[\beta\left(\Lambda^{AA}+h\right)\right]+\exp\left(-\frac{\beta J^{AA}_{z}}{4}\right)\cosh\left(\frac{\beta J^{AA}_{\perp}}{2}\right)}
\label{eq7}
\\
\tanh\left[\frac{\beta}{2}\left(\Lambda^{A}+h\right)\right]&=&\frac{\exp\left(\frac{\beta J^{AB}_{z}}{4}\right)\sinh\left[\beta\left(\Lambda^{AB}+h\right)\right]+ \exp\left(-\frac{\beta J^{AB}_{z}}{4}\right)\sinh\left(\frac{\beta \Delta^{AB}}{2}\right)}{\exp\left(\frac{\beta J^{AB}_{z}}{4}\right)\cosh\left[\beta\left(\Lambda^{AB}+h\right)\right]+\exp\left(-\frac{\beta J^{AB}_{z}}{4}\right)\cosh\left(\frac{\beta \Delta^{AB}}{2}\right)}
\label{eq8}
\\
\tanh\left[\frac{\beta}{2}\left(\Lambda^{B}+h\right)\right]&=&\frac{\exp\left(\frac{\beta J^{BB}_{z}}{4}\right)\sinh\left[\beta\left(\Lambda^{BB}+h\right)\right]}{\exp\left(\frac{\beta J^{BB}_{z}}{4}\right)\cosh\left[\beta\left(\Lambda^{BB}+h\right)\right]+\exp\left(-\frac{\beta J^{BB}_{z}}{4}\right)\cosh\left(\frac{\beta J^{BB}_{\perp}}{2}\right)}
\label{eq9}
\\
\tanh\left[\frac{\beta}{2}\left(\Lambda^{B}+h\right)\right]&=&\frac{\exp\left(\frac{\beta J^{AB}_{z}}{4}\right)\sinh\left[\beta\left(\Lambda^{AB}+h\right)\right]- \exp\left(-\frac{\beta J^{AB}_{z}}{4}\right)\sinh\left(\frac{\beta \Delta^{AB}}{2}\right)}{\exp\left(\frac{\beta J^{AB}_{z}}{4}\right)\cosh\left[\beta\left(\Lambda^{AB}+h\right)\right]+\exp\left(-\frac{\beta J^{AB}_{z}}{4}\right)\cosh\left(\frac{\beta \Delta^{AB}}{2}\right)}
\label{eq10}.
\end{eqnarray}
The knowledge of the values of the variational parameters $\lambda^{\gamma \delta}$ allows for a complete determination of Gibbs energy as a function of the external field $h$ and temperature $T$. On this basis all further thermodynamic properties of the magnetic bilayer can be calculated.\\ 

\subsection{Planar magnetizations, compensation and phase transition temperature}

The total magnetization per site can be obtained from the thermodynamic formula:
\begin{equation}
m=-\left(\frac{\partial G}{\partial h}\right)_{T}
\label{eq11}.
\end{equation}
This leads to the expression:
\begin{equation}
m=\frac{1}{2}\left(m_A+pm_B\right)
\label{eq12}
\end{equation}
where $m_A$ and $m_B$ are the planar magnetizations in the plane $A$ and $B$ (for occupied sites), respectively. They are given by the formula:
\begin{equation}
m_{\gamma}=\frac{1}{2}\tanh\left[\frac{\beta}{2}\left(\Lambda^{\gamma}+h\right)\right] 
\label{eq13}
\end{equation}
where $\gamma = A,B$.
Let us remind that the fields $\Lambda^{\gamma}$ in eq.(\ref{eq13}) are related to the parameters $\lambda^{\gamma\delta}$ by equations (\ref{eq7}-\ref{eq10}).\\
For the antiferromagnetic interaction $J^{AB}_{z}$ between the planes $A$ and $B$, magnetizations $m_A$ and $m_B$ have opposite signs. The compensation temperature $T_{\rm comp}$ can be found from the condition:
\begin{equation}
m_A=-pm_B
\label{eq14}
\end{equation}
(for $m_A,m_B\neq 0$), which yields $T_{\rm comp}\leq T_{\rm C}$, where $T_{\rm C}$ is the phase transition temperature.\\
In order to determine the critical temperature of continuous phase transitions $T_{\rm C}$, we put $h=0$ and $\lambda^{\gamma \delta} \to 0$ in the variational equations (\ref{eq7}-\ref{eq10}). As a consequence, these equations transform into a linear homogeneous set, which is solvable when:
\begin{equation}
{\rm det}\, \hat{M}=0.
\label{eq15}
\end{equation}
The matrix $\hat{M}$ has the following form:

\begin{equation}
\hat{M}=
\left[ \begin{array}{cccc}
2\left(3-2C^{AA}\right) & 0 & p\left(2-C^{AA}\right) & 0 \\
0 & 4p\left(2-C^{AB}\right) & -C^{AB} & 0 \\
0 & 2\left(4p-1-2pC^{BB}\right) & 0 & 2-C^{BB} \\
4\left(2-C^{AB}\right) & 0 & \left(p-1\right)\left(2-C^{AB}\right) & -C^{AB} 
\end{array} \right].
\label{eq16}
\end{equation}
The temperature-dependent coefficients $C^{\gamma \delta}$ in eq.(\ref{eq16}) are given by the formulas:
\begin{equation}
C^{\gamma \gamma}=1+\exp\left(-\frac{\beta_{\rm C} J^{\gamma \gamma}_{z}}{2}\right)
\cosh\left(\frac{\beta_{\rm C} J^{\gamma \gamma}_{\perp}}{2}\right)
\label{eq17}
\end{equation}
($\gamma=A,B$) and
\begin{equation}
C^{AB}=1+\exp\left(-\frac{\beta_{\rm C} J^{AB}_{z}}{2}\right),
\label{eq18}
\end{equation}
where $\beta_{\rm C}=1/k_{\rm B}T_{\rm C}$. Therefore, eq.(\ref{eq15}) is the equation for the phase transition temperature. This equation has no general analytic solution, so that the critical temperatures can only be found numerically.\\
The numerical calculations based on the above mentioned formalism will be presented in the next section.\\

\section{Numerical results and discussion}

In order to study the phase diagrams and magnetization behaviour of the bilayer, the numerical calculations have been performed. The main physical problem concerns antiferromagnetic interactions between two non-equivalent subsystems. For some Hamiltonian parameters the existence of compensation temperatures has been evidenced in Figs.2 and 3. Fig.2 is prepared for the case when intraplanar interactions in layers $A$ and $B$ are of ising type. In turn, in Fig.3, the isotropic Heisenberg interplanar interaction is present in both planes. In both figures the upper curves, with filled symbols, correspond to the phase transition temperatures, whereas the lower ones (open symbols) denote the compensation point temperatures. In Fig.2, $p^{\star}$ stands for the concentrations in plane $B$ for which the compensation temperature becomes equal to the phase transition temperature. Thus, $p^{\star}$ is a characteristic concentration below which, for the given Hamiltonian parameters, the compensation phenomenon does not take place.\\

\begin{figure}
\includegraphics[scale=0.25]{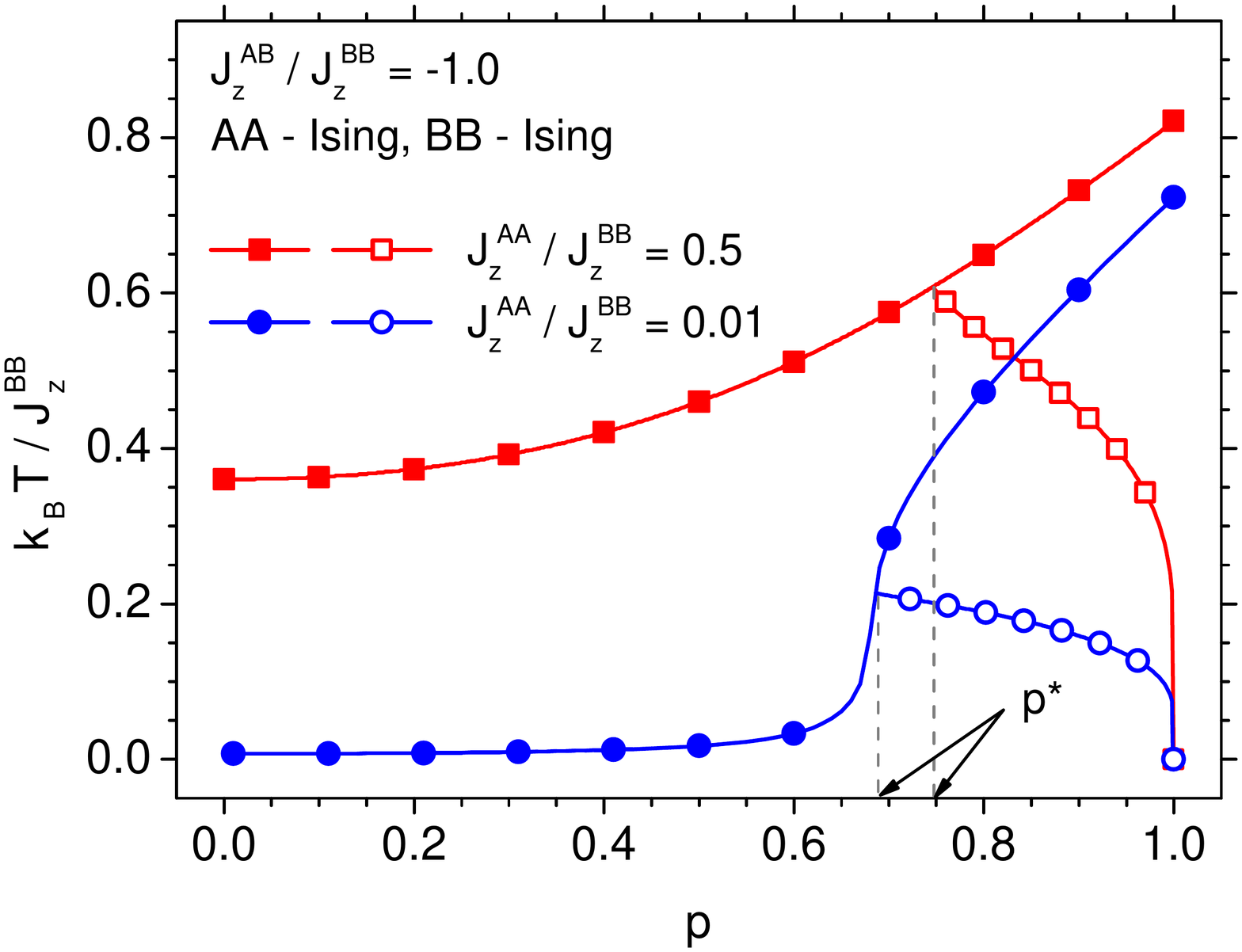}
\caption{\label{fig:fig2}The phase transition temperatures $T_{\rm C}$ (filled symbols) and the compensation temperatures $T_{\rm comp}$ (empty symbols) vs. concentration $p$. Both planes are of Ising-type. By $p^{\star}$ we denote the characteristic concentration for which $T_{\rm comp}=T_{\rm C}$.}
\end{figure}

\begin{figure}
\includegraphics[scale=0.25]{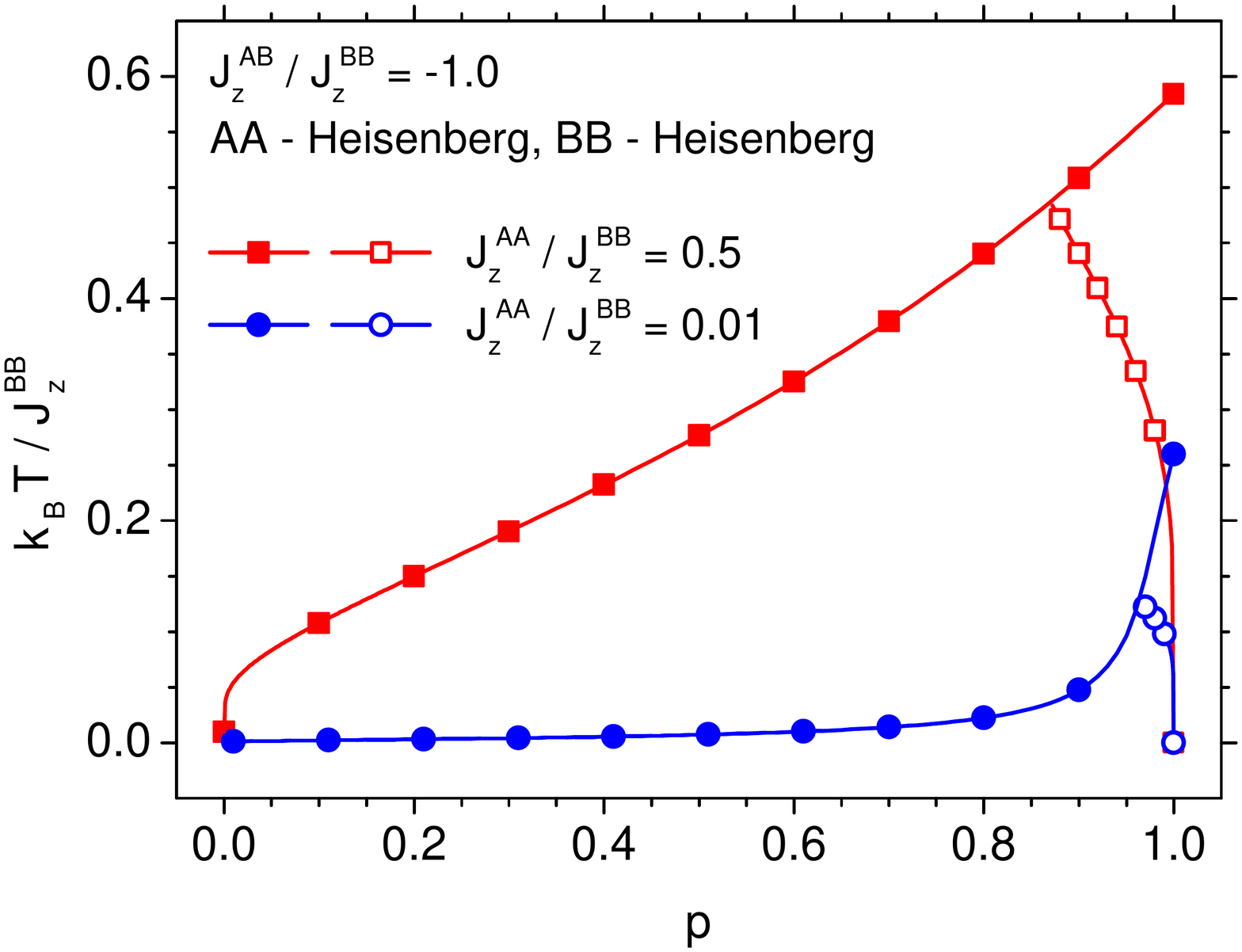}
\caption{\label{fig:fig3}The phase transition temperatures $T_{\rm C}$ (filled symbols) and the compensation temperatures $T_{\rm comp}$ (empty symbols) vs. concentration $p$. Both planes are of Heisenberg-type.}
\end{figure}

It can be noted that for the Ising model (Fig.2), $p^{\star}$ is higher for the case with $J_{z}^{AA}/J_{z}^{BB}=0.5$ than for the case when $J_{z}^{AA}/J_{z}^{BB}=0.01$. It means that when the interaction within plane $A$ becomes weaker, the concentration  $p^{\star}$ decreases. The opposite tendency can be seen in Fig.3. i.e., for the Heisenberg intraplanar couplings. Moreover, for concentration $p \to 0$, i.e., when magnetic atoms in plane $B$ are absent, the remaining plane $A$ reveals ordered phase for the Ising model (Fig.2), but for the Heisenberg interactions in that plane (Fig.3) the phase transition temperature goes to zero. We note that the vanishing of magnetism for the isotropic Heisenberg 2D model is in agreement with the rigorous Mermin-Wagner theorem \cite{Mermin1}. It is seen in Fig.3 that when the $A$-plane is magnetically weak ($J_z^{AA}/J_z^{BB}=0.01$) the phase transition  temperature is significant only when $p \to 1$ in plane $B$. Let us observe that for $p=1$, the compensation take place at $T=0$, since in that case the ground state magnetizations of both planes are equal to $1/2$ and opposite in sign.\\

\begin{figure}
\includegraphics[scale=0.25]{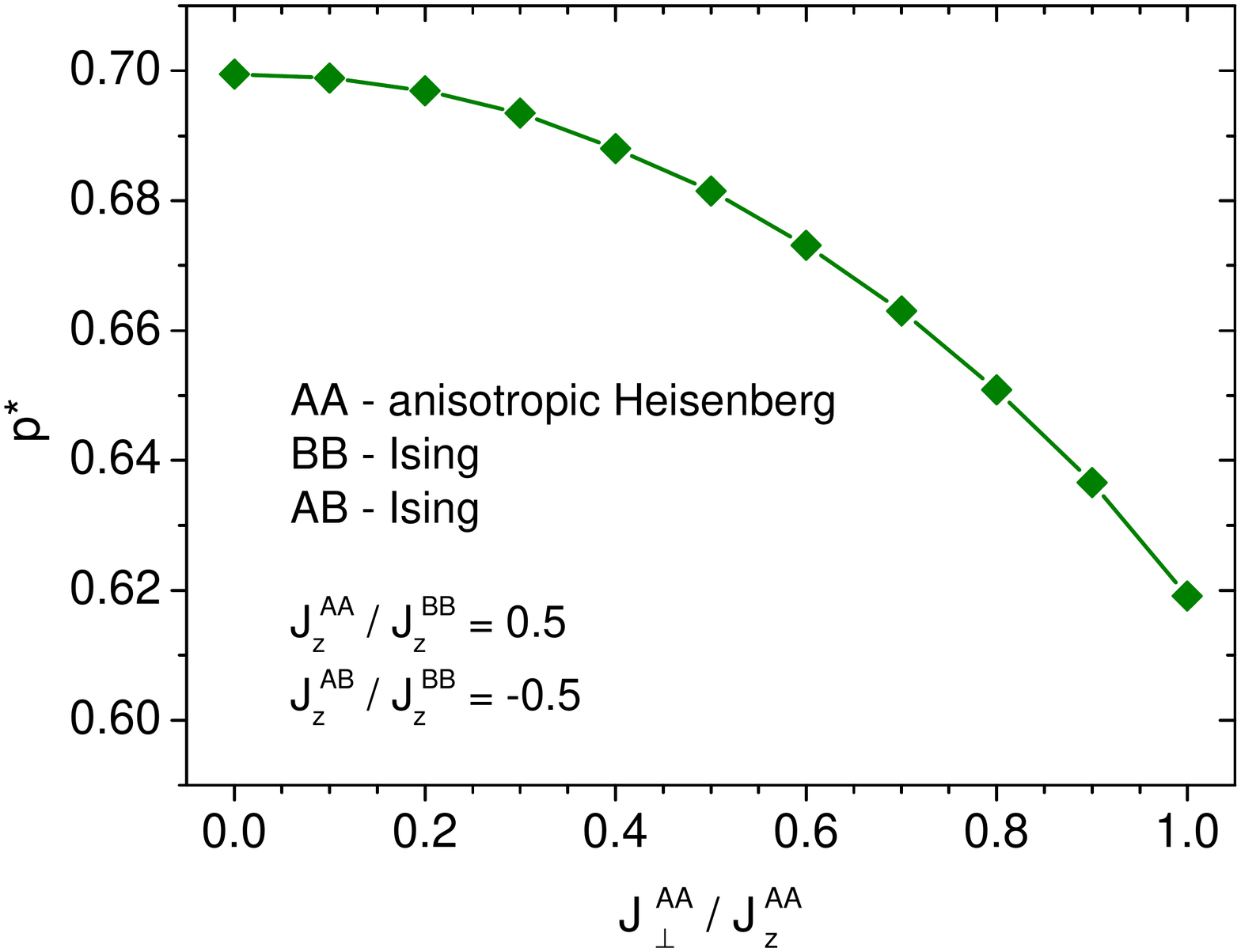}
\caption{\label{fig:fig4}The characteristic concentration $p^{\star}$ vs. anisotropy parameter $J_{\perp}^{AA}/J_{z}^{AA}$ when the plane $A$ is of anisotropic Heisenberg-type and the plane $B$ is of Ising-type.}
\end{figure}

\begin{figure}
\includegraphics[scale=0.25]{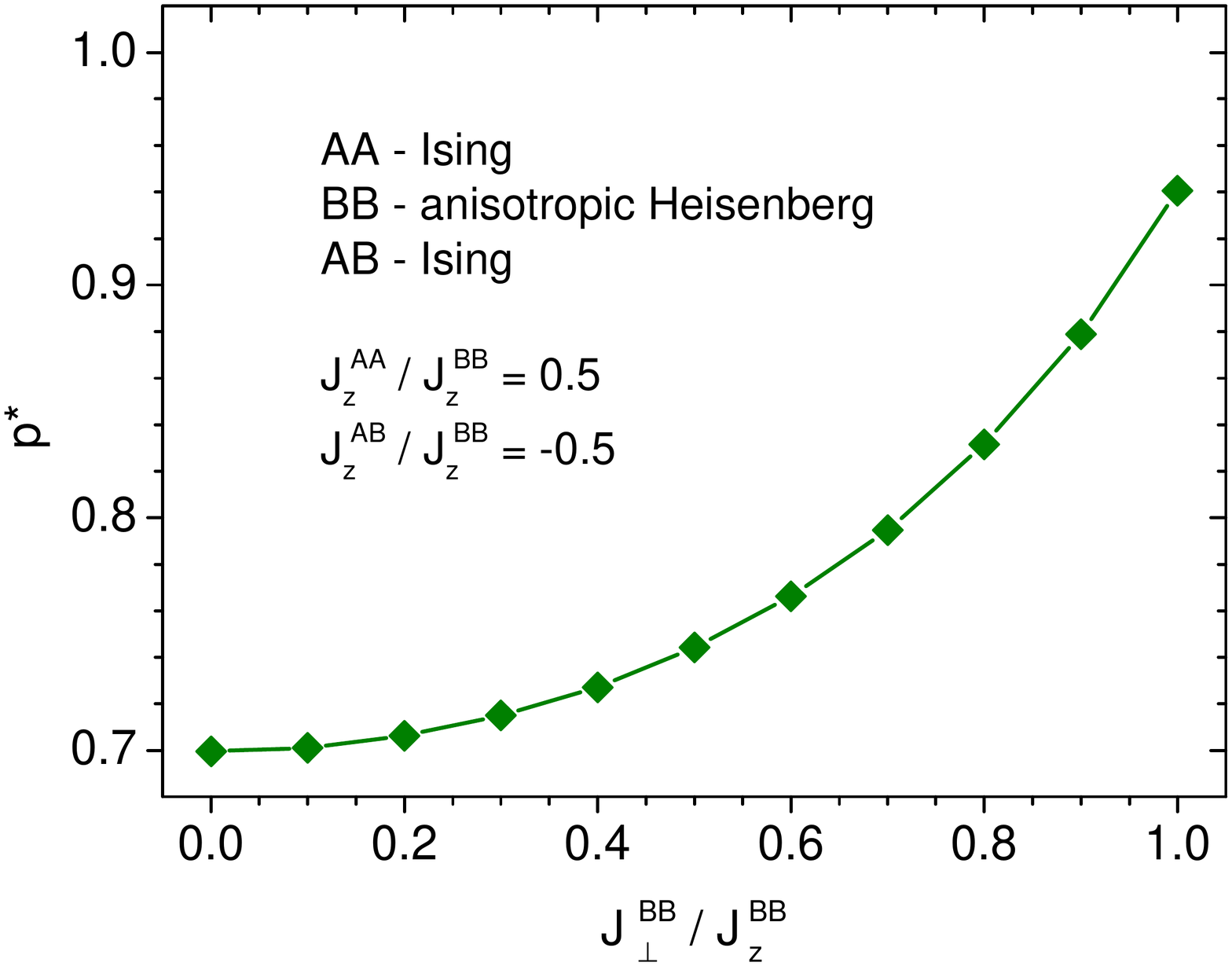}
\caption{\label{fig:fig5}The characteristic concentration $p^{\star}$ vs. anisotropy parameter $J_{\perp}^{BB}/J_{z}^{BB}$ when the plane $A$ is of Ising-type and the plane $B$ is of anisotropic Heisenberg-type.}
\end{figure}

In order to study the behaviour of characteristic concentration $p^{\star}$ in more detail, we plot $p^{\star}$ as a function of anisotropy parameters $J_{\perp}^{AA}/J_{z}^{AA}$ (in Fig.4) and $J_{\perp}^{BB}/J_{z}^{BB}$ (in Fig.5). The whole range of variability of the anisotropy parameter is presented in both figures, starting from the zero value, when both planes are of Ising type, till the value of 1, when the intraplanar interaction within either plane $A$ (Fig.4) or plane $B$ (Fig.5) is fully isotropic. Comparing Figs.4 and 5 one can see that the behaviour of $p^{\star}$ vs. anisotropy is very different in both cases. In Fig.4 a decrease of $p^{\star}$ vs. $J_{\perp}^{AA}$ is observed, while in Fig.5 $p^{\star}$ increases with increase of $J_{\perp}^{BB}$. One should notice that both planes $A$ and $B$ are magnetically non-equivalent: the exchange interaction $J_{z}^{AA}$ in plane $A$ is half of the value $J_{z}^{BB}$ in plane $B$. On the basis of Figs.4 and 5 one can conclude that the increase or decrease of $p^{\star}$ as a function of anisotropy depends on which plane is anisotropic in such bilayer system. In both cases, the physical reason for the shift of $p^{\star}$ point is the weakening of magnetization in the Heisenberg plane when the anisotropy decreases.\\

\begin{figure}
\includegraphics[scale=0.25]{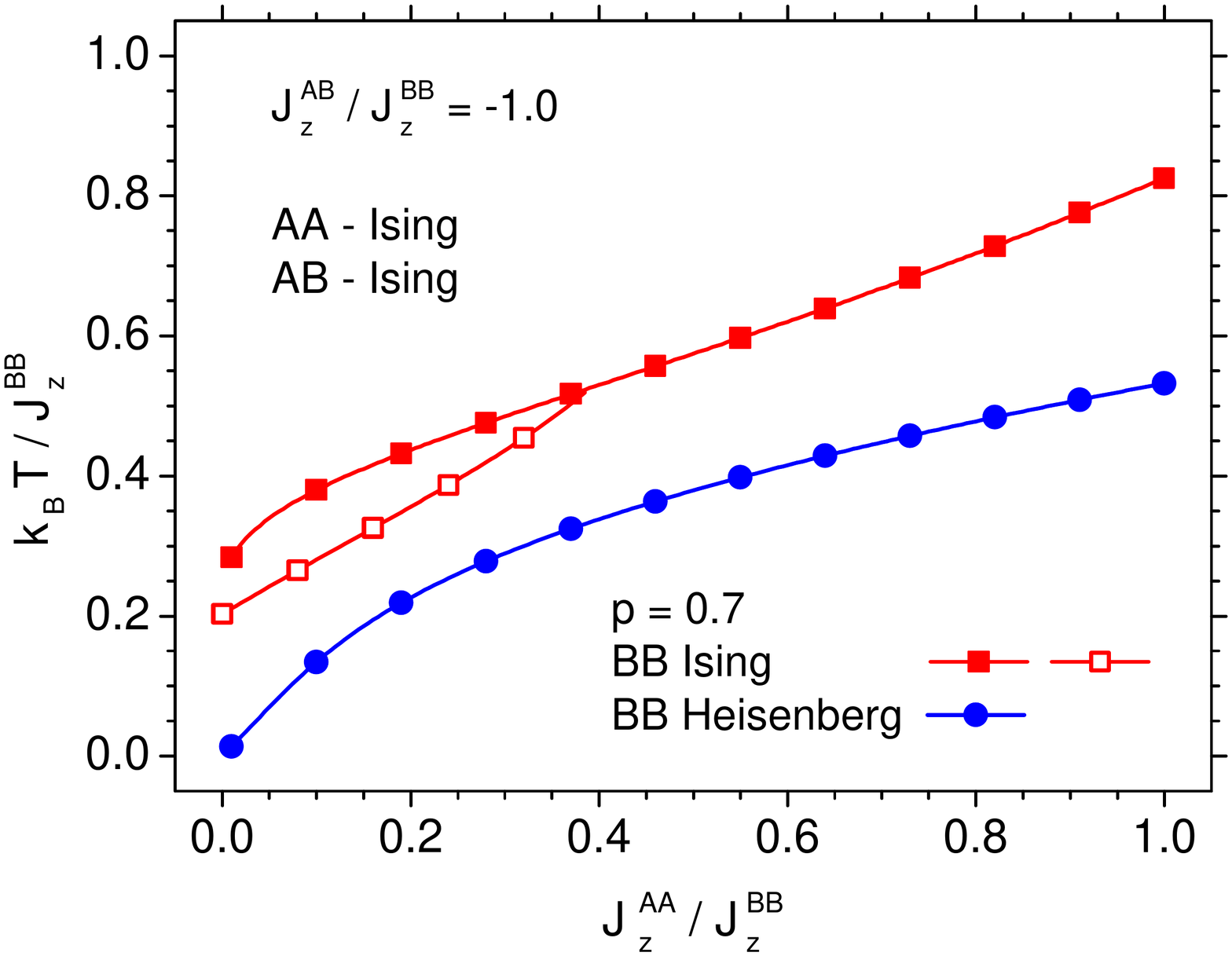}
\caption{\label{fig:fig6}The phase transition and compensation temperatures vs. asymmetry parameter $J_{z}^{AA}/J_{z}^{BB}$. The interplanar interaction is $J_{z}^{AB}/J_{z}^{BB}=-1$ and the concentration amounts to $p=0.7$.}
\end{figure}

\begin{figure}
\includegraphics[scale=0.25]{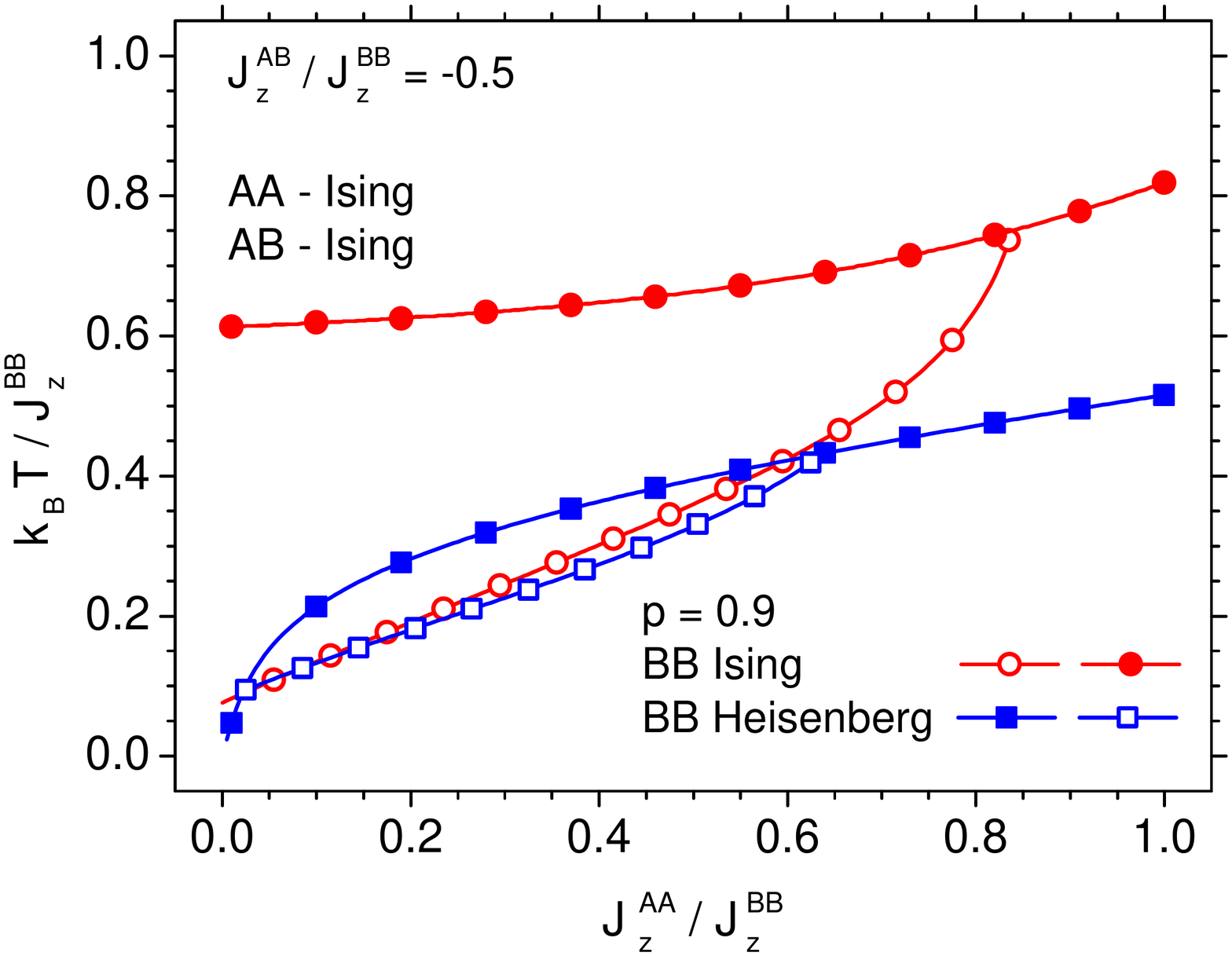}
\caption{\label{fig:fig7}The phase transition and compensation temperatures vs. asymmetry parameter $J_{z}^{AA}/J_{z}^{BB}$. The interplanar interaction is $J_{z}^{AB}/J_{z}^{BB}=-0.5$ and the concentration amounts to $p=0.9$.}
\end{figure}

The fact that the planes $A$ and $B$ are magnetically non-equivalent is of important consequences, as it can be seen in Figs.6 and 7. In both these figures the horizontal axis corresponds to the ratio  $J_{z}^{AA}/J_{z}^{BB}$. In Fig.6 the concentration $p=0.7$ is assumed together with relatively strong interlayer antiferromagnetic coupling $J_{z}^{AB}/J_{z}^{BB}=-1$. When the plane $B$ is of Ising type, the compensation effect takes place only for the low values of $J_{z}^{AA}$, i.e., when the plane $A$ is magnetically weak in comparison with the plane $B$. However, when the plane $B$ itself becomes magnetically weaker, as is in the Heisenberg case (lower curve), the compensation effect may not occur at all. That results from the fact that the magnetization in the plane $B$ is lower than in the plane $A$ for all the temperatures $T<T_C$. This conclusion can be confirmed on the basis of Fig.7. This figure is made for higher concentration, $p=0.9$, but for weaker interlayer coupling  $J_{z}^{AB}/J_{z}^{BB}=-0.5$. For such a choice of parameters the compensation effect is present in both cases, i.e., when the plane $B$ is either of Ising- or  Heisenberg type. However, for the Heisenberg case the range where the effect occurs is evidently narrower.\\

\begin{figure}
\includegraphics[scale=0.25]{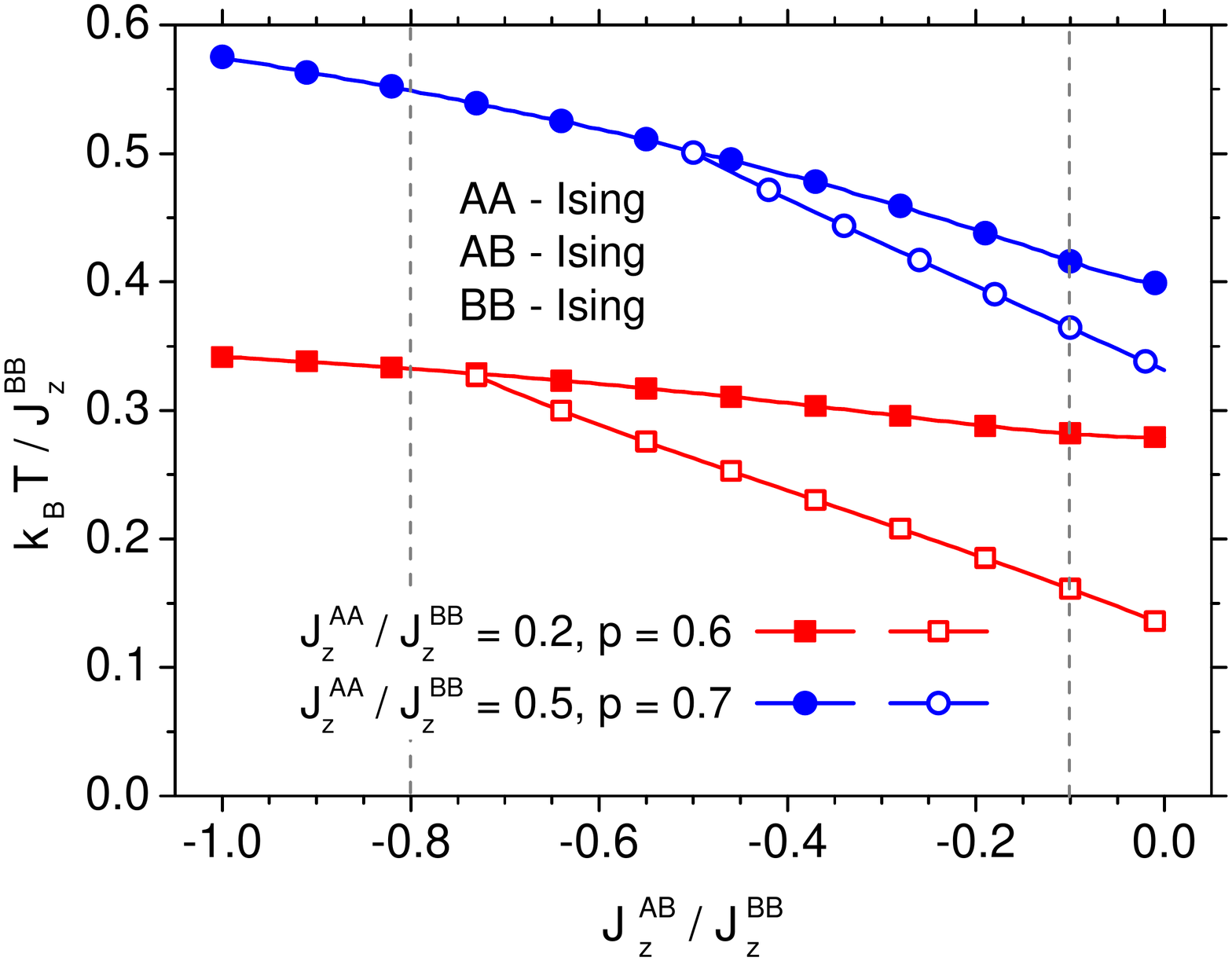}
\caption{\label{fig:fig8}The phase transition and compensation temperatures vs. interplanar interaction parameter $J_{z}^{AB}/J_{z}^{BB}$. All interactions are of Ising-type. The vertical dashed lines mark exemplary cross-sections for magnetization calculations (see Figs.9 and 10).}
\end{figure}

In Fig.8 we present the dependence of the phase transition temperature and compensation temperature on the interlayer coupling strength $J_{z}^{AB}$. It turns out that a weak antiferromagnetic coupling favours the compensation effect although, at the same time, it reduces the phase transition temperature. As is seen in that figure, the phenomenon depends on the concentration $p$ and the asymmetry of couplings $J_{z}^{AA}/J_{z}^{BB}$. The two vertical dashed lines in Fig.8 mark the temperature cross-sections of the diagram for which the magnetization calculations are presented in subsequent Figs. 9 and 10.\\ 

\begin{figure}
\includegraphics[scale=0.25]{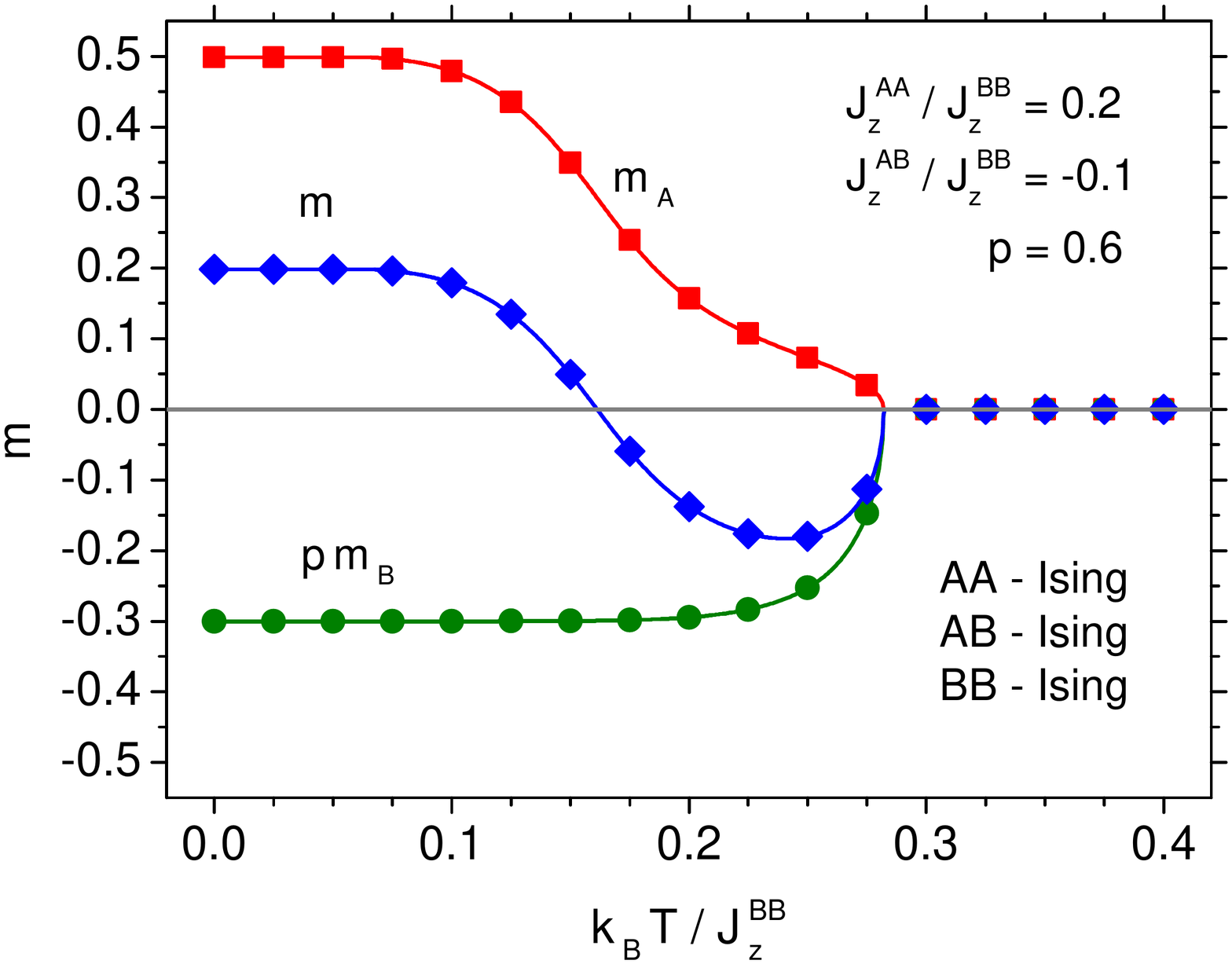}
\caption{\label{fig:fig9}The planar magnetizations $m_{A}$, $pm_{B}$ and the total magnetization $m$ vs. temperature for 
$J_{z}^{AA}/J_{z}^{BB}=0.2$, $J_{z}^{AB}/J_{z}^{BB}=-0.1$ and concentration $p=0.6$. The total magnetization shows compensation point for $m=0$ and $T_{\rm comp}\leq T_{\rm C}$.}
\end{figure}

\begin{figure}
\includegraphics[scale=0.25]{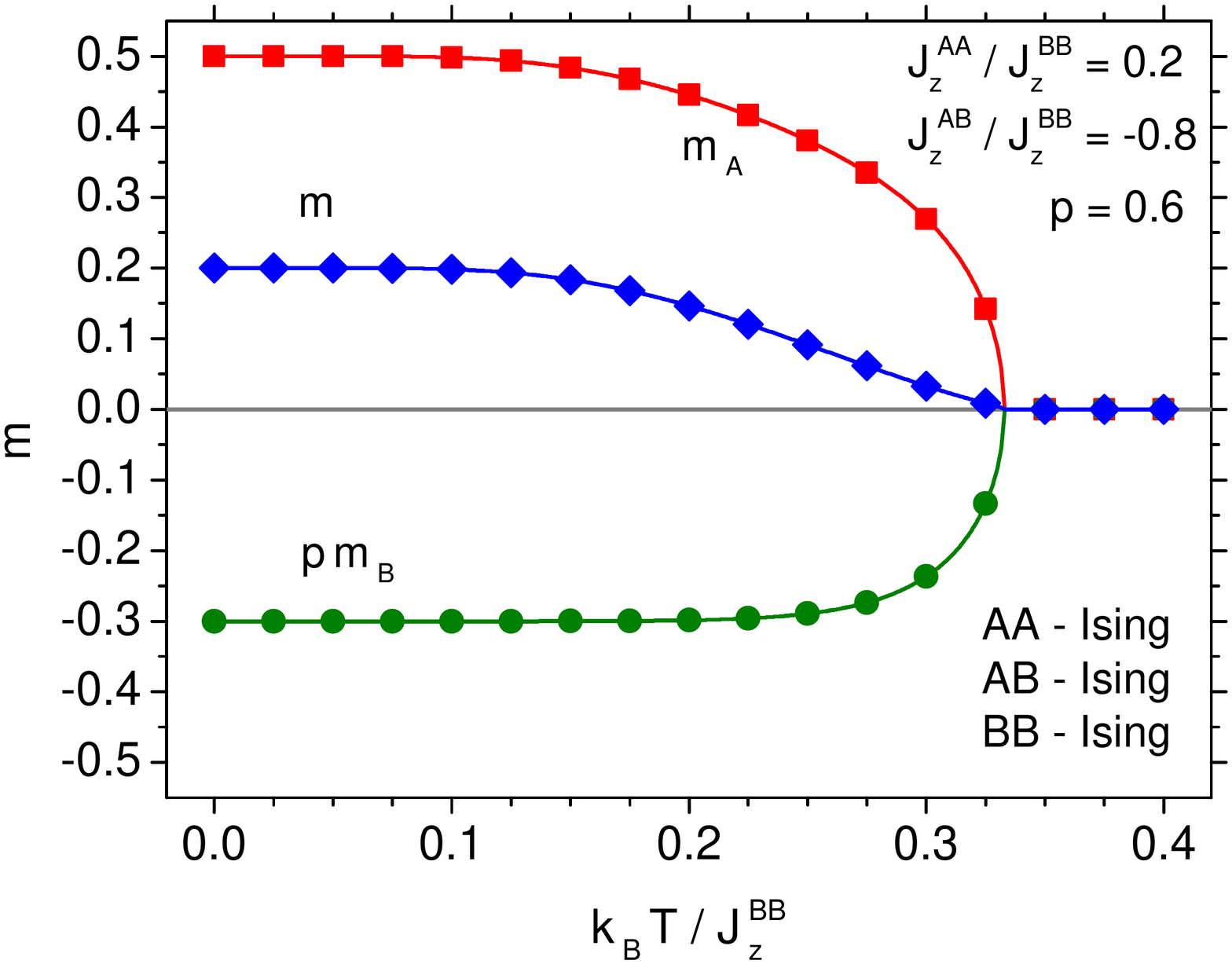}
\caption{\label{fig:fig10}The planar magnetizations $m_{A}$, $pm_{B}$ and the total magnetization $m$ vs. temperature for 
$J_{z}^{AA}/J_{z}^{BB}=0.2$, $J_{z}^{AB}/J_{z}^{BB}=-0.8$ and concentration $p=0.6$. In this case the compensation effect doesn't occur.}
\end{figure}

 For instance, in Fig.9 one can see the dependence of plane magnetizations  $m_{A}$ and $m_{B}$ vs. temperature for $p=0.6$, $J_{z}^{AA}/J_{z}^{BB}=0.2$ and $J_{z}^{AB}/J_{z}^{BB}=-0.1$. The total magnetization $m=m_A+pm_B$ is also drawn showing the compensation effect for $m=0$. The physical reason for the compensation phenomenon in the asymmetric bilayer is as follows: The plane $A$, which is not diluted, has the maximum magnetization $m_A=1/2$ at $T=0$. Since the interaction in this plane is weak, the magnetization curve quickly drops when the temperature increases. On the other hand, the magnetization curve of layer $B$ with strong intraplanar coupling has quasi-rectangular shape. However, this magnetization is lower than $m_A$ and tends to the maximum value of $p/2$ when $T \to 0$. In consequence, the magnetization for plane $B$ has the absolute value of $|pm_B|$, which is lower than $m_A$ for low temperatures, but higher than $m_A$ in the high temperature region. This leads to the compensation effect at some intermediate temperature. Such picture also explains why the shift of compensation point depends on the concentration $p$ in the plane $B$ and on the ratio of interaction parameters in both planes. For instance, when the shape of magnetization curve $m_A$ becomes nearly rectangular in temperature and the magnetization $pm_B$ is weak, the compensation effect may completely vanish, as can be seen in Fig.10. This figure has been created for $J_{z}^{AB}/J_{z}^{BB}=-0.8$, which corresponds to the left vertical line in Fig.8. For stronger interactions $J_z^{AA}$ in the plane $A$ and higher concentration $p$ in the plane $B$, the phase transition temperature increases, as can be expected. At the same time, due to more pronounced role of intraplanar interactions the magnetization in both planes becomes stronger, thus reducing the effect of asymmetry. In the case presented in Fig.10 the magnetization $m_A$ is larger than $|pm_B|$ for all temperatures $T<T_{\rm C}$.\\

\begin{figure}
\includegraphics[scale=0.25]{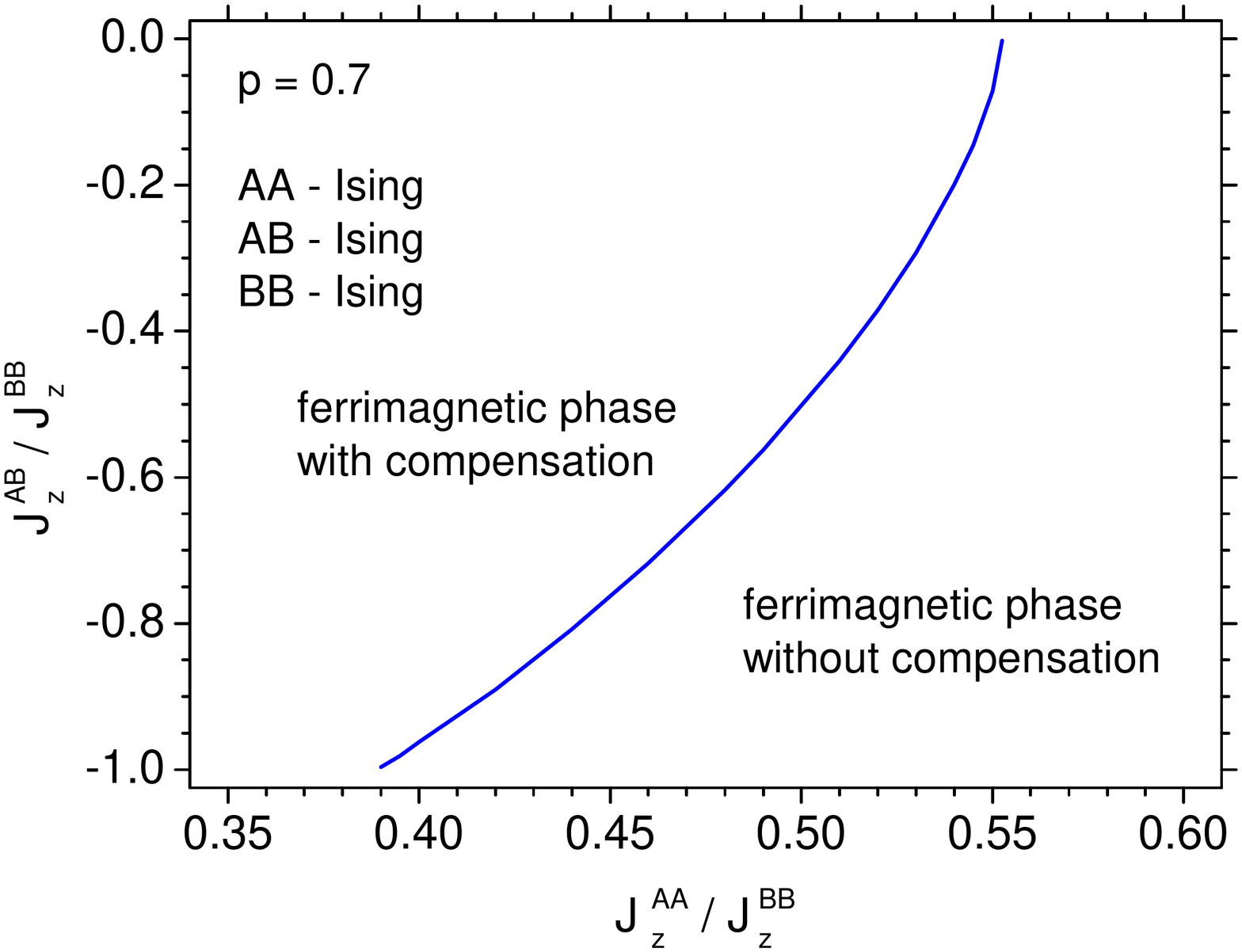}
\caption{\label{fig:fig11}The phase diagram for $p=0.7$, with all interactions of Ising type.}
\end{figure}

\begin{figure}
\includegraphics[scale=0.25]{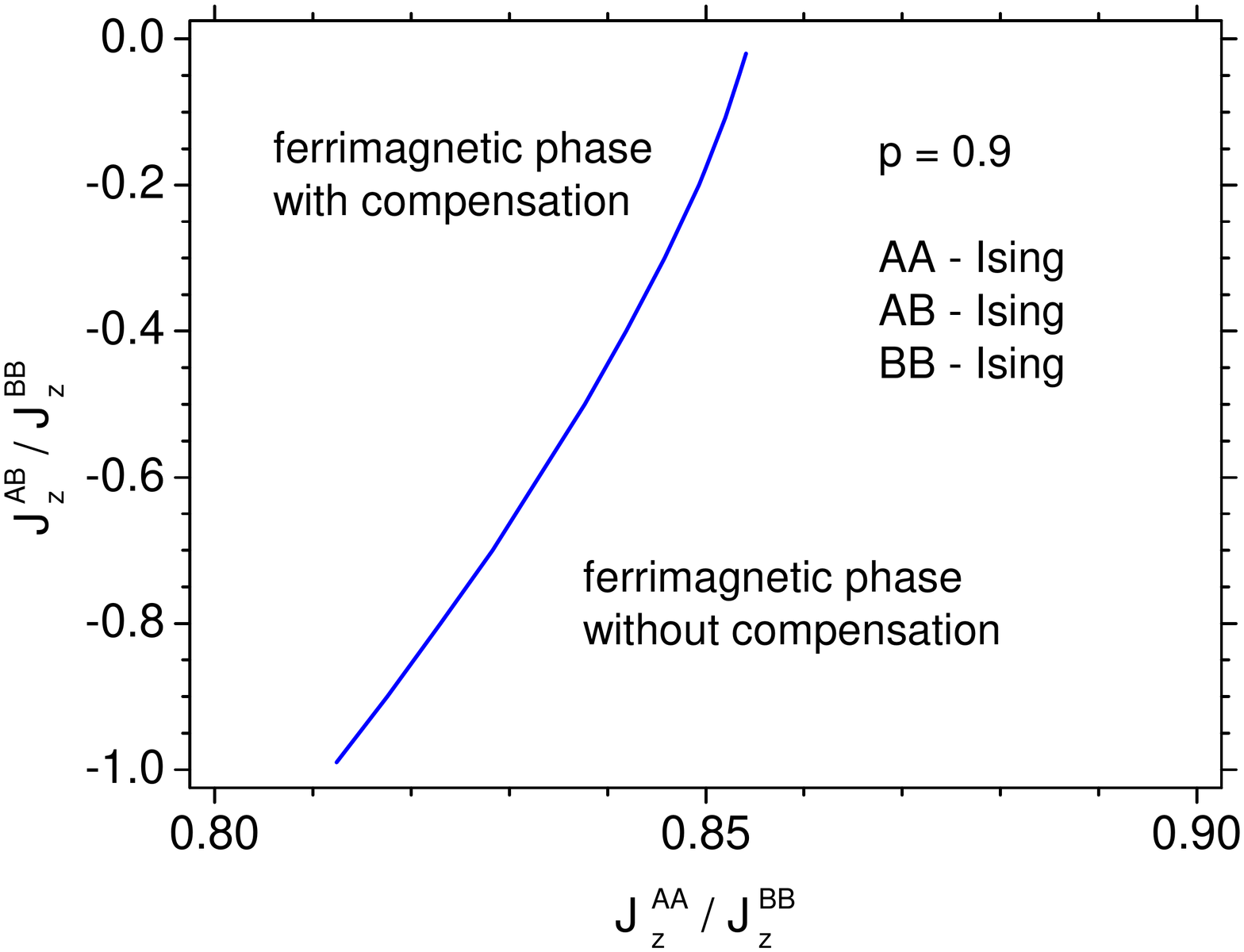}
\caption{\label{fig:fig12}The phase diagram for $p=0.9$, with all interactions of Ising type.}
\end{figure}

As it follows from the results presented above, the subspace of model parameters within the range of our interest can be divided into two areas. One corresponds to the existence of a ferrimagnetic phase for which compensation phenomenon is absent at any temperature. The other involves ferrimagnetic phase for which compensation takes place at certain temperature $T_{comp}$. Therefore, it can be useful to plot some phase diagrams showing the areas corresponding to the existence of both phases. For this purpose we present Fig.11, in which a critical line separating both kinds of ferrimagnetic phases is plotted for $p=0.7$ as dependent on interplanar coupling $J_{z}^{AB}/J_{z}^{BB}$ and intraplanar coupling in plane A $J_{z}^{AA}/J_{z}^{BB}$, for all interactions of Ising type. The diagram confirms that compensation is robust for weaker interplanar coupling and for more pronounced intraplanar coupling asymmetry. An analogous phase diagram for $p=0.9$ is illustrated in Fig.12. It is evident that for higher concentration of magnetic component in the plane $B$, the border between both phases is almost vertical and the critical value of interplanar coupling only slightly depends on the intraplanar interactions in the plane $A$. Moreover, the range occupied by ferrimagnetic phase without compensation is significantly reduced.

\section{Conclusions}

The Cluster Variational Method in Pair Approximation has been applied to the asymmetric Heisenberg-Ising bilayer with antiferromagnetic interaction between the ferromagnetic planes. The method allows to perform the complete characterization of the system thermodynamics, based on the knowledge of the Gibbs free energy. The asymmetry of the system was introduced by different interaction parameters in both planes and by site dilution of one plane ($B$). In the case when the diluted plane $B$ is characterized by stronger exchange interactions than the plane $A$, the compensation effect can occur. The value of characteristic concentration $p^{\star}$, below which no compensation takes place, depends on the Hamiltonian parameters and is very different for the Ising and Heisenberg models. The behaviour of the compensation temperature can be explained by analysis of magnetization curves, which change their shape for different anisotropy and concentration parameters. In particular, when two planes are non-interacting (for $J_{z}^{AB}=0$), the Mermin-Wagner theorem is fulfilled \cite{Mermin1}. In our opinion, the model of asymmetric bilayer is more general and closer to experimental situation than the ideal symmetric one. From one side, it allows for interaction of the bilayer with the substrate; on the other side, it takes into account the non-magnetic impurities and surface disorder.\\

The effect of compensation temperature obtained here is robust and we believe that it cannot be just an artifact of the PA method. The low-temperature magnetization is dominated by the ordering of the undiluted plane, so that the orientation of the total magnetization follows the magnetization direction in plane $A$. On the other hand, the intraplanar couplings in plane $A$ are weaker than in plane $B$, so that in higher temperatures $m_A$ drops significantly and the plane $A$ gets only magnetized by the influence of the plane $B$ with stronger intraplanar couplings. Therefore, the total magnetization near the critical temperature is rather governed by the magnetization in $B$ plane and follows its direction. Since the signs of $m_A$ and $m_B$ are opposite, an occurrence of a compensation point can be expected for some intermediate temperature separating both regimes, on condition that the concentration of the magnetic component in plane $B$ is sufficiently high, but not equal to $p=1$. Nevertheless, confirmation of emergence of compensation behaviour by more sophisticated approaches and/or Monte Carlo simulations would be of great value.\\

This model can be extended for other low-dimensional systems, for instance, magnetic stripes, also including the case of higher spins ($S>1/2$).
We think that the studies can also be extended for multilayers consisting of the antiferromagnetic planes which are magnetically asymmetric and diluted.\\

\section{Acknowledgements}

This work has been supported by the Polish Ministry of Science and Higher Education on a special purpose grant to fund the research and development activities and tasks associated with them, serving the development of young
scientists and doctoral students.

\newpage

\bibliographystyle{elsarticle-num}








\end{document}